\documentclass[prl,twocolumn,superscriptaddress,showpacs,preprintnumbers,amssymb]{revtex4}
\usepackage{graphicx}

\begin{document}
\preprint{LA-UR-02-3874}

\thispagestyle{myheadings} \markright{{\em LA-UR-02-1265}}

 \title{Experimental Evidence for a Glass forming ``Stripe Liquid"
 in the Magnetic Ground State of \mbox{La$_{1.65}$Eu$_{0.2}$Sr$_{0.15}$CuO$_{4}$}}

 \author{B. Simovi\v c}
 \affiliation{Condensed Matter and Thermal Physics,
              Los Alamos National Laboratory, Los Alamos, NM 87545}
 \author{P. C. Hammel}
 \affiliation{Condensed Matter and Thermal Physics,
              Los Alamos National Laboratory, Los Alamos, NM 87545}
\author{M. H\"{u}cker}
 \affiliation{Physics Department Brookhaven National
Laboratory, Upton, New-York 11973}
 \author{B. B\"uchner}
 \affiliation{RWTH Aachen, Aachen Germany}
 \author{A. Revcolevschi}
 \affiliation{Laboratoire de Chimie des Solides, Universit\'{e} Paris-Sud, 91405, Orsay Cedex, France}

 \begin{abstract}
We report measurements of the longitudinal ($^{139}T_1^{-1}$) and
transverse ($^{139}T_2^{-1}$) decay rates of the magnetization of
$^{139}$La nuclei performed in a high quality single crystal of
La$_{1.65}$Eu$_{0.2}$Sr$_{0.15}$CuO$_{4}$. We observe a dramatic
slowing of the Cu 3d spins manifested as a sharp increase of both
$^{139}T_1^{-1}$ and $^{139}T_2^{-1}$ below 30 K. We find that in
this temperature range the fluctuations involve a unique time
scale $\tau $ which diverges as $(T-T_{\rm A})^{-1.9}$  with
$T_{\rm A}\thickapprox 5$ K. This behavior is distinct from the
 continuous freezing observed in underdoped La$_{1-x}$Sr$_x$CuO$_4$ which involves
a distribution of energy barriers. By contrast, in
La$_{1.65}$Eu$_{0.2}$Sr$_{0.15}$CuO$_{4}$, the freezing below 30K
is intrinsic to its magnetic ground state and the observed power
law supports the existence of a glass forming ``charge stripe
liquid".
\end{abstract}
 \pacs{ 71.27.+a, 71.45.Lr, 74.25.Ha, 74.72.Dn,75.60.Ch, 76.60.Es }

\maketitle

Charge inhomogeneities in doped transition metal-oxides have
aroused great interest because of their possible implication in
superconductivity. Indeed, it has been
proposed~\cite{ZaanenJ:HigsSd,ZaanenJ:SupSod,KivelsonSA:Elelpd,EmeryVJ:Spipem}
that the superconducting state in underdoped cuprates may consist
of a charge stripe phase, in which hole-rich fluctuating stripes
act as antiphase boundaries between undoped antiferromagnetic (AF)
domains. Experimentally, the presence of a stripe order was
inferred from elastic neutron scattering
measurements~\cite{TranquadaJM:Eviscs} in compounds that are not
superconducting: doped lanthanum nickelates and the rare earth
co-doped lanthanum cuprate
La$_{1.48}$Nd$_{0.4}$Sr$_{0.12}$CuO$_{4}$. In the latter, the
substitution for La ions by isovalent rare earth ions (Nd or Eu)
induces a structural transition from the low temperature
orthorhombic (LTO) to the low temperature tetragonal (LTT) phase
in which a magnetic ground state occurs in lieu of the
superconducting state~\cite{BuchnerB:Cribds}.

In both structures, LTO and LTT, the CuO$_6$ octahedra are tilted
inducing a staggered buckling of the CuO$_2$ plane
~\cite{AxeJD:Strilc}. The difference between the two phases
consists only of  a rotation of the tilt axis from $[110]_{\rm
HTT}$ in the LTO phase to  $[100]_{\rm HTT}$ or $[010]_{\rm HTT}$
alternating along the c axis in the LTT phase. Though specific to
lanthanum cuprates, the LTT structure was
shown~\cite{BuchnerB:Cribds, KlaussH-H:antosm} to have the
remarkable property of generating either a magnetic or
superconducting ground state with similar critical temperature at
fixed optimal Sr$^{2+}$ doping (x=0.15) depending only on the
amplitude of the buckling, controlled by the Eu or Nd
concentration. This suggests the same physics underlies the two
outcomes, but little is known about the fundamental mechanism that
connects them. In this context, it is important to gain further
insight into the low energy properties of the magnetic ground
state in rare earth co-doped lanthanum cuprates.

In this Letter, we report NMR measurements performed on $^{63}$Cu
and $^{139}$La nuclei in a
La$_{1.65}$Eu$_{0.2}$Sr$_{0.15}$CuO$_{4}$ single crystal. For the
first time, we investigate  the slowing down of the Cu 3d spins in
this compound by measuring two relaxation rates: $^{139}T_1^{-1}$
and $^{139}T_2^{-1}$. For a given orientation of the crystal in a
static field $\mathbf H_0$, $T_1^{-1}$ probes the spectral weight
of the transverse fluctuations in the hyperfine field $H_{\rm hf}$
{\em only} at the Larmor frequency $\omega_0$ of the nuclear
spins. However, because of the contribution to the hyperfine
interaction that commutes with the Zeeman Hamiltonian (secular
term), $T_2^{-1}$ also measures the time scale $\tau$ of the
longitudinal fluctuations at frequencies down to $\omega_{\rm hf}
= \gamma H_{\rm hf}$~\cite{Slichter}, where $\gamma$ is the
gyromagnetic ratio of the given nucleus.

It is now well established that upon cooling, the fluctuations of
the Cu 3d spins exhibit glassy spin freezing~\cite{CurroNJ:Inhlfs,
JulienMH:Chascs}, however the mechanism~\cite{ek:glass,
SchmalianJ:Strgs-, WestfahlH:Sel-ge, Preprint-Grousson} underlying
this behavior remains poorly understood. Our $^{139}T_2^{-1}$
measurements have afforded a key new insight into these slow spin
dynamics revealing the glassy dynamics of the ``stripe liquid'' as
the underlying cause of the observed spin dynamics in
La$_{1.65}$Eu$_{0.2}$Sr$_{0.15}$CuO$_{4}$. Below 30 K we observe
that the time constant $\tau$ describing the electron spin
dynamics responsible for the $^{139}$La spin relaxation begins to
increase very sharply with decreasing temperature following the
power law $\tau \sim (T-T_{\rm A})^{-\alpha}$ over three decades
in frequency where 5 K$\, \lesssim T_{\rm A} \lesssim 6.5$ K and
$\alpha =1.9 \pm 0.1$. This temperature dependence of the time
scale of the fluctuations is consistent with the behavior
predicted using the dynamical mode-coupling (MC) approximation for
a glass forming charge stripe liquid~\cite{Preprint-Grousson,
SchmalianJ:Strgs-, WestfahlH:Sel-ge}. The dynamics of relaxation
is determined  by the time scale $\tau$ and the stretching
exponent $\beta(T)$ of the time correlation function $G(t)=\exp
[-(t/\tau)^{\beta}]$~\cite{Gotze}. We will show that the observed
power law for $\tau (T)$ together with the temperature dependence
of the exponent $\beta$ given  in the MC regime of  glass forming
liquids~\cite{Gotze}, can account for the peculiar temperature
dependence of $^{139}T_1^{-1}$ below 30 K.

The single 5mm$^{3}$ crystal used in this study was grown using
the traveling solvent floating zone method under oxygen pressure
of 3 bar \cite{Ammerahl:TZF}. Diffraction data indicates a sharp
structural transition at 135$\pm$2 K in agreement with NMR, and
dc magnetization measurements.  The $^{139}$La($I = 7/2$) and
$^{63}$Cu ($I = 3/2$) NMR measurements were made on the central
($m_I = +\frac{1}{2} \leftrightarrow -\frac{1}{2})$ transition.
$^{139}T_1^{-1}$ and $^{63}T_1^{-1}$ were measured by monitoring
the recovery of the magnetization after an inversion pulse in a
field of 84 kOe for Cu, and 60 and 75.2 kOe for La.
$^{139}T_1^{-1}$ was measured at 75.2 kOe and 73.8 kOe by
monitoring the amplitude of the spin echo as a function of the
delay between $\pi/2$ and $\pi$ radio-frequency pulses.

\begin{figure}[tb]
 \centering
\includegraphics[width=\linewidth]{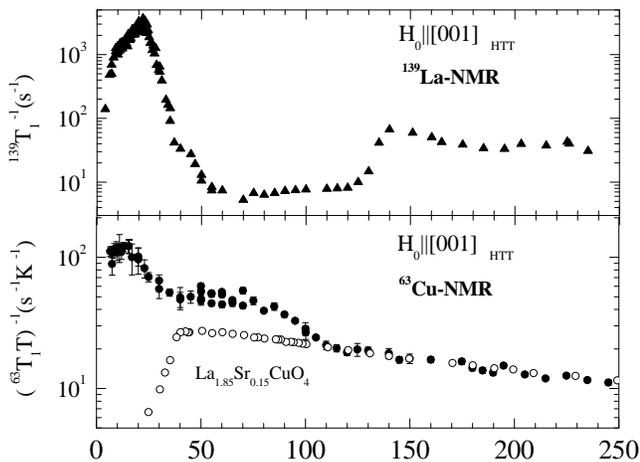}
\caption{(a) \(^{139}T_1^{-1}\) versus T (b) $^{63}(T_1T)^{-1}$
versus T for La$_{1.65}$Eu$_{0.2}$Sr$_{0.15}$CuO$_{4}$ (solid
circles) and La$_{1.85}$Sr$_{0.15}$CuO$_4$ (open circles; from
Ref.~\onlinecite{OhsugiS:CuNsts})} \label{1}
\end{figure}

We show in Fig.~1 the overall temperature dependence
of~$^{139}T_1^{-1}$ (Fig.~1a) and $^{63}(T_{1}T)^{-1}$(Fig.~1b).
Since Cu and La nuclei are located in and outside the CuO$_2$
plane respectively, they couple differently to fluctuations.
$^{139}$La is weakly coupled to the electronic properties in the
plane and its nuclear spin lattice relaxation rate
\(^{139}T_1^{-1}\) is therefore a very sensitive probe of the
lattice dynamics that cause fluctuations of the electric field
gradient (EFG) at the La site. We find indeed that above 50 K, the
magnetization recovery curves are well fitted by the standard
expression~\cite{SuterA:Sepqmc} for a pure quadrupole relaxation
with a single $T_1$. In Fig.~1a, the sharp drop of
\(^{139}T_1^{-1}\) at 135 K clearly reflects the first order
structural transition from the LTO to the LTT phase. Below this
transition, no particular features are observed down to 50K. At
lower temperature, \(^{139}T_1^{-1}\) increases dramatically to
reach a broad maximum around 20 K  with concomitant changes in the
shape of the recovery curves. These sudden changes in the
relaxation clearly manifest qualitative changes in the mechanism
of relaxation of the magnetization of  $^{139}\rm La$ nuclei. We
fit the recovery curves below 50 K to the expression
\(M(t)=M_{0}[1-2 \exp (-\sqrt{t/T_1})]\)\cite{footnote}

In contrast to $^{139}$La, the $^{63}$Cu nuclear spin is strongly
coupled to the spin of the single hole localized on the 3d$_{\rm
x^{2}-y^{2}}$ orbital. Thus, for a static field $\mathbf
H_{0}\parallel [001]_{\rm HTT}$, the quantity $^{63}(T_1T)^{-1}$
reflects the weighted {\bf q} average of $\chi " ({\mathbf q},
\omega_{0})$, the imaginary part of the in-plane dynamical spin
susceptibility at the frequency $\omega_{0}$. The magnetization
recovery curves for $^{63}$Cu were therefore fitted to the
standard expression~\cite{SuterA:Sepqmc} for a purely magnetic
relaxation over the entire temperature range investigated assuming
a single $T_1$. The temperature dependence of $^{63}(T_1T)^{-1}$
is shown in Fig.~1b. No sharp discontinuity is observed at 135 K,
and $^{63}(T_1T)^{-1}$ follows the temperature dependence observed
in the normal state of
La$_{1.65}$Sr$_{0.15}$CuO$_{4}$.~\cite{OhsugiS:CuNsts} However, we
see in Fig.~1b that a significant deviation from this canonical
behavior occurs below 100 K: $^{63}(T_1T)^{-1}$ exhibits a two
step increase toward a maximum around 16 K indicating a two step
freezing of the Cu 3d spins.  The interpretation of the
temperature dependence of $^{63}(T_1T)^{-1}$ is however hampered
by the dramatic suppression of the Cu signal that occurs below
100K. As previously shown~\cite{CurroNJ:Inhlfs}, this phenomena
bears witness to a strong decrease of the relaxation time
$^{63}T_2$, indicating significant slowing of the Cu 3d spins
fluctuations in this temperature range. The reason for this sudden
change is however very elusive. It is worth pointing out that
below 70 K, a resistivity upturn occurs~\cite{HuckerM:Consot} in
this compound, suggesting a possible connection with charge
inhomogeneities. In order to better understand how charge
inhomogeneities and the slowing of Cu 3d spins are related, we
performed a detailed study of the temperature dependence of the
time scale of these fluctuations by means of \(^{139}T_2^{-1}\)
measurements.

\begin{figure}[tb]
 \centering
\includegraphics[width=0.9\linewidth]{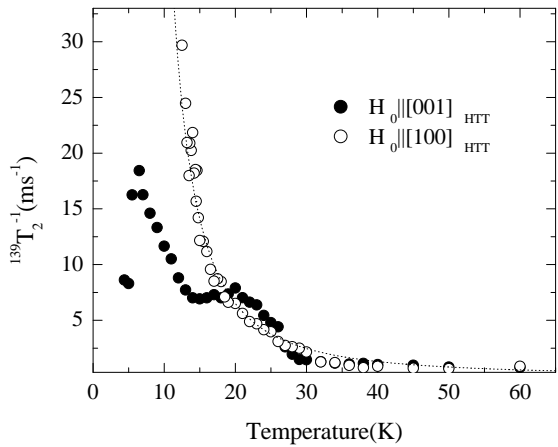}
\caption{ \(^{139}T_2^{-1}\) versus $T$. The decays of ${M}_\perp$
are exponentials for ${\mathbf H_0}
\parallel [100]_{\rm HTT}$ whereas for
$\mathbf H_0 \parallel [001]_{\rm HTT})$, a stretched relaxation
is observed  below 30 K. For consistency, we fit all the decays to
a single exponential (see text). Dotted line: $1.16$ $(T-T_{\rm
A})^{-1.9}$ with $T_{\rm A}=5$ K} \label{2}
\end{figure}

The relaxation rate $T_1^{-1}$ is  expected to be maximum when the
characteristic time $\tau$  of the fluctuations becomes comparable
to $\omega_0^{-1}$. However, because of the secular part of the
hyperfine interaction, the transverse nuclear magnetization
$M_\perp$ decays rapidly to zero: the Larmor precession rate of
nuclear spins is shifted by random hopping of $H_{\rm hf}$ along
$\mathbf H_0$. If \( \tau > \omega_{0}^{-1}$, $H_{\rm hf} \) is
not averaged out and \(T_2^{-1}\) will no longer be proportional
to \(T_1^{-1}\) because of the contribution of the secular term
$\gamma^2 H^2_{{\rm hf}} \tau$ to $T_2^{-1}$\cite{Slichter}. In
this regime, \(T_2^{-1}\)(T) will follow the temperature
dependence of $\tau$ until the fluctuations  become slower than \(
\omega_{\rm hf}=\gamma H_{\rm hf} \)\cite{Slichter}.

In Fig.~2, we show \(^{139}T_2^{-1}\) as a function of temperature
for $\mathbf H_0$ along [001] and [100] respectively. For $\mathbf
H_0$ along [001], \(^{139}T_2^{-1}\) goes through a first maximum
around 20 K indicating that $\tau \sim \omega_{0}^{-1}$ at this
temperature. However, below this, and in sharp contrast with
\(^{139}T_1^{-1}\) reported on Fig.~1a, the rate
\(^{139}T_2^{-1}\) continues to increase gradually, reaching a
second maximum at $6.5\pm 0.5$ K. In this temperature range, the
decay of ${M}_\perp$ is caused by fluctuations that are too slow
to relax the nuclear magnetization. The narrow maximum of
\(^{139}T_2^{-1}\) at 6.5 K  followed by a sharp drop below,
indicates that, for this temperature, we have reached  the
motional narrowing limit i.e., $\tau \approx \omega_{\rm
hf,z}^{-1}$. We can thus conclude that at 6.5 K, \( \tau \approx
{^{139}T_2}= 50\, \mu \)s (i.e., $\omega_{\rm hf, z} \approx 20
\rm kHz$) and hence that $H_{\rm hf, z}=31$ G at the La site.

For $\mathbf H_0 \parallel$ [100], a glance at Fig.~2 shows that
\(^{139}T_2^{-1}\) increases  continuously, the NMR signal
disappearing rapidly below 10 K. The absence of any broad maximum
for $\tau \sim \omega_0^{-1}$ indicates that for this orientation,
\(^{139}T_2^{-1}\) is dominated by the secular contribution:
$\gamma^2 H^2_{{\rm hf}, x} \tau$. This can be easily understood
on the sole basis of the strong anisotropy of the hyperfine field
at La site. Indeed, at 20 K we have $\tau^{-1} =
\omega_{0}=T_{2}\gamma^2 H^2_{{\rm hf, x}}$ and we thus estimate
$H_{{\rm hf},x} \approx 948$ G. Note that the values of $H_{{\rm
hf},z}$ and $H_{{\rm hf},x}$ deduced from our experiment are
remarkably close to the 43 G and 990 G deduced from La-NQR in the
AF magnetic ordered  state of the parent compound $\rm
La_2CuO_{4+\delta}$ \cite{NishiharaH:NQRNs1}. This also confirms
the magnetic origin of the relaxation in this temperature range.

It is widely accepted that the freezing to a clusters spin glass
state observed in underdoped $\rm La_{2-x}Sr_xCuO_4$ results from
a continuous relaxation slowing down from the high temperature
metallic phase\cite{MarkiewiczRS:Clus-g}. The time scale of the
fluctuations obeys the Vogel-Fulcher relaxation law:
$\tau(T)=\tau_{0} e^{E_{\rm a}/(T-T_{\rm g})}$ with
$\tau_{0}\approx 0.03 \rm ps$ and $T_{\rm g}$ is the glass
transition.~\cite{MarkiewiczRS:Clus-g} The energy barriers $E_{\rm
a}$ are spatially distributed and in a wide range of doping
\(^{139}T_1^{-1}\)(T) can be easily modelled at low T assuming a
gaussian distribution centered on $E_{0} \approx 50 \rm K$ and of
width $ \Delta \approx 80 \rm K $.~\cite{CurroNJ:Inhlfs} Based on
\(^{139}T_1^{-1}\)(T) measured in
La$_{1.65}$Eu$_{0.2}$Sr$_{0.15}$CuO$_{4}$ (different sample from
the present study) it was concluded that the freezing observed in
the magnetic ground state of this compound is similar to
underdoped $\rm La_{2-x}Sr_xCuO_4$.~\cite{CurroNJ:Inhlfs} However,
the maximum of \(^{139}T_1^{-1}\) reported in
ref\cite{CurroNJ:Inhlfs} is observed at 12K whereas ZF-$\mu \rm
SR$ which frequency scale is $\sim 1 \rm MHz$  detects the onset
of static moment around 25K\cite{KlaussH-H:antosm} in agreement
with the present work. Also, the non monotonic increase of the
freezing temperature upon doping detected by ZF-$\mu \rm SR$ in
$\rm La_{1.8-x}Eu_{0.2}Sr_xCuO_4$\cite{KlaussH-H:antosm} argues
against doping independent freezing properties in rare earth
co-doped lanthanum cuprates.

On Fig.3 we show the overall temperature dependence below 50K of
$\tau$ deduced from \(^{139}T_2^{-1}\) (Fig.3a) and
\(^{139}T_1^{-1}\)(Fig.3b). We first analyze our data assuming the
above mentioned  Vogel-Fulcher dynamics. Based on
\(^{139}T_1^{-1}\) we found $E_{0}=230 \pm 20 \rm K$ and $\Delta =
80K \pm 10 \rm K$ and the fit (dotted line) is superimposed to the
data  on  Fig.3b. We also plot $\tau(T)$ expected for the energy
barrier $E_{0}=230 \rm K$ on Fig.3a (dotted line). The inadequacy
of this approach is evident. \newline In the interval 30-10K, the
measured $\tau(T)$ can be fit to an Arrhenius law assuming a
single energy barrier $E_{0}$ or a power law \((T-T_{\rm
A})^{-\alpha} \). The two fits are shown on Fig.3a and we found on
one hand $E_{0}=52 \pm 1 \rm K$ and $\tau_{0} = 1.3 \pm 0.01 \rm
ns$ and on the other hand $\alpha =1.9 \pm 0.1$ and 5 K $\leqslant
T_{\rm A}\leqslant 6.5$K. We see on Fig.3 that the Arrhenius law
completely fails to account for \(^{139}T_1^{-1}\)(T). However,
the power law has been recently predicted to occur in the MC
regime of a glass forming ``stripe liquid"
\cite{Preprint-Grousson}. In this model, the stripe phase is in a
thermal equilibrium liquid state above $T_{\rm A}$, with defects
in the stripe pattern able to wander freely in the CuO$_2$ plane
\cite{WestfahlH:Sel-ge}. Approaching $T_{\rm A}$ \cite{Gotze}, the
dynamics is determined by two temperature dependent parameters: a
{\em unique} time scale $\tau$ following a power law dependence on
$T$, and the stretched exponent $\beta$, through the homogeneous
time correlation function \(G(t)=\exp[-(t/\tau)^{\beta}]\). As
discussed in Ref.~\onlinecite{Gotze}, \(\beta =-\ln 2/\ln
(1-\lambda)\) with \(\lambda=\mu (T-T_{\rm A})/T_{\rm A}\) and
$\mu$, a parameter. On the temperature range where the power law
is observed,  it thus allows only two free parameters to describe
\(^{139}T_1^{-1}\)(T): (i) the temperature $T_{\rm A}$ which
should be below 6.5 K according to Fig.~3a;  and (ii) the
parameter $\mu$ which is set by the temperature at which $\beta$
departs significantly from 1 (i.e., where $\lambda$ departs from
1/2). In describing $T_1(T)$ the choice of $\mu$ determines the
position of the maximum of \(^{139}T_1^{-1}\) below 30 K while the
value of $T_{\rm A}$ fixes the width of the peak. We find that the
experimental temperature dependence of \(^{139}T_1^{-1}\) reported
in Fig.~3b is well described for parameter values $T_{\rm A}=5$ K
and $\mu =0.1389$ (corresponding to a temperature of 25 K at which
$G(t)$ departs from simple exponential behavior). We find
excellent agreement between the theory and the data for
\(^{139}T_1^{-1}\) as shown in Fig.~3b; particularly striking is
its ability to describe the peculiar steep downturn of
\(^{139}T_1^{-1}\) around 5 K.

\begin{figure*}[t]
\centering
\includegraphics[width=0.9\linewidth]{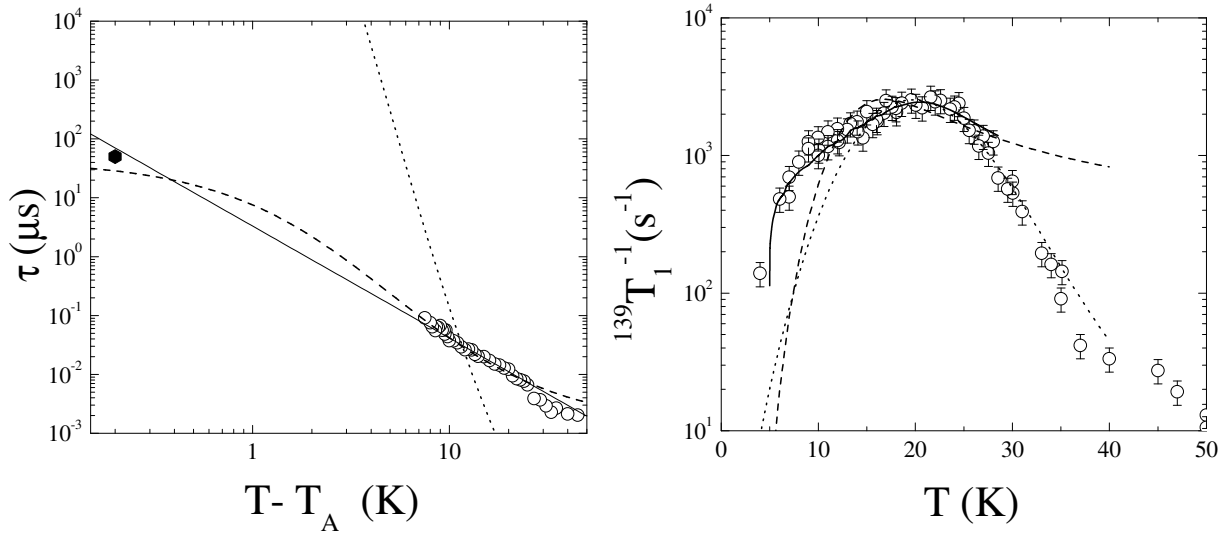}
\caption{(a) $\tau$(T) deduced from \(^{139}T_2^{-1}\) data
reported in Fig.~2. (b) \(^{139}T_1^{-1}\)(T) below 50K. On both
panels we compare three different models for the dynamics. Solid
line: the power law $\tau(T) = 3.3(T-T_{\rm A})^{-1.9}$ with
$T_{\rm A}=5$K. The decay of the homogeneous time correlation
function approaching $T_{\rm A}$ is then determined by the
quantity $\lambda(T)=1-\mu (T-T_{\rm A})/T_{\rm A}$ with $\mu =
0.1389$ and $T_{\rm A}=5\rm K$. \(^{139}T_1^{-1}\) (T) is
calculated by Fourier transformation of the stretched correlation
function (see in text). Dashed line: Arrhenius behavior $\tau(T)
=\tau_{0} \rm Exp(-E_{0}/T)$ with a single energy barrier
$E_{0}=52\pm 1 \rm K$ and $\tau_{0}=1.3 \pm 0.01 \rm ns$ and
\(^{139}T_1^{-1}\) (T) $ \propto \tau/(1+(\omega_{0}\tau)^{2})$.
Dotted line: gaussian distribution of energy barriers with
$\tau_{0}$ fixed to 0.03 ps as in
Ref.~\onlinecite{CurroNJ:Inhlfs}.  For this fit $E_{0}=230 \pm 20
\rm K$ and a distribution width $\Delta =80\pm 10$K.}
\end{figure*}

To conclude, we investigated the spin freezing in
La$_{1.65}$Eu$_{0.2}$Sr$_{0.15}$CuO$_{4}$ by measuring
independently the temperature dependence of two key quantities:
(i) the time scale of the wandering motion of the electron moment
fluctuations (ii) the temperature dependence of the spectral
weight of the very same fluctuations. Our data  reveal a dramatic
slowing down below 30K which cannot be extrapolated to the high
temperature metallic phase but appears to be {\em intrinsic} to
the  magnetic ground state.  Detailed analysis shows that the
freezing is well described by a power law indicating a
characteristic temperature around 5 K.  Such a power law is
generally expected to occur in the MC regime of glass forming
liquids, the temperature $T_{\rm A}$ marking a cross-over to a
Vogel-Fulcher dynamics approaching $T_{\rm g}$. For the particular
case of the charge stripe liquid state, $T_{\rm g}$ is predicted
to be only 10$\%$ smaller than the temperature $T_{\rm
A}$~\cite{WestfahlH:Sel-ge} indicating the predominance of the
power law over a wide temperature range. So, in view of the
present work, the underlying cause for the observed spin freezing
in La$_{1.65}$Eu$_{0.2}$Sr$_{0.15}$CuO$_{4}$ is likely to be the
evolution of a charge-stripe liquid toward a glass as predicted in
ref\cite{WestfahlH:Sel-ge, Preprint-Grousson}. \\
The authors thank M. Grousson, J. Schmalian and P. G. Wolynes for
valuable discussions. The work at Los Alamos National Laboratory
and Brookhaven were performed under the auspices of the US
Department of  Energy (Contract No. DE-AC02-98CH10886).

\bibliographystyle{prsty}
\bibliography{rer}

\begin{thebibliography}{10}
\bibitem{footnote}
This choice is arbitrary
since the recovery curve of the magnetization of $^{139}$La
nuclear spins is  multi-exponential due to the contribution of the
non-saturated satellite transitions to the
relaxation\cite{SuterA:Sepqmc}. The fit to a stretched exponential
with a stretching exponent of 0.5 has been previously proposed in
ref\cite{CurroNJ:Inhlfs} to account for the fact that the recovery
cannot be fit to  the standard expression for a pure magnetic
relaxation assuming a single $T_1$.
\end{thebibliography}
\end{document}